\documentclass[prl,aps, babel,twocolumn,showpacs]{revtex4-1}
\usepackage{amsmath,amssymb,amstext,amscd,amsthm,amsopn,graphicx,indentfirst,mathrsfs}

\usepackage{psfrag}

\begin{document}

\title{Scaling Properties of Weak Chaos in Nonlinear Disordered Lattices}
\author{Arkady Pikovsky}
\affiliation{\mbox{Department of Physics and Astronomy, Potsdam University,
  Karl-Liebknecht-Str 24, D-14476, Potsdam, Germany}}
\author{Shmuel Fishman}
\affiliation{\mbox{Department of Physics, Technion Israel Institute of Technology, 32000 Haifa, Israel}}

\date{\today}
%\date{August 24, 2007}

\begin{abstract}
The Discrete Nonlinear Schroedinger Equation with a random potential in one dimension is studied as a dynamical system. It is characterized by the length, the strength of the random potential and by the field density that determines the  effect of nonlinearity. The probability of the system to be regular is established numerically and found to be a scaling function. This property is used to calculate the asymptotic properties of the system in regimes beyond our computational power.
\end{abstract}

%05.45.-a Nonlinear dynamics and chaos
%72.15.Rn Localization effects (Anderson or weak localization)
%05.50.+q   Lattice theory and statistics (Ising, Potts, etc.)
%(see also 64.60.Cn Order–disorder transformations, and 75.10.Hk Classical spin models)
%63.20.Pw Localized modes
%63.20.Ry Anharmonic lattice modes
%63.70.+h Statistical mechanics of lattice vibrations and displacive phase transitions
%03.75.Kk Dynamic properties of condensates; collective and hydrodynamic
%    excitations, superfluid flow
%05.30.Jp Boson systems (for static and dynamic properties of Bose-Einstein
%    condensates, see 03.75.Hh and 03.75.Kk)
%05.60.Gg Quantum transport
%73.20.Fz Weak or Anderson localization
%63.50.-x Vibrational states in disordered systems
%03.75.Nt Other Bose-Einstein condensation phenomena
%%%PACS from Ref.20
%03.75.Kk Dynamic properties of condensates; collective and hydrodynamic
%    excitations, superfluid flow
%03.75.Nt Other Bose-Einstein condensation phenomena
%05.60.Gg Quantum transport
%%%PACS from Ref.21 B.Shapiro
%03.75.Kk Dynamic properties of condensates; collective and hydrodynamic
%    excitations, superfluid flow
%71.55.Jv Disordered structures; amorphous and glassy solids

\maketitle

The  Nonlinear Schr\"odinger Equation (NLSE), in absence of a
potential, in the continuum is an integrable
problem~\cite{Sulem-99}. It is relevant for the description of
Bose-Einstein condensates (where it is known as the
Gross-Pitaevskii Equation~\cite{Dalfovo-99,
*Pitaevskii-Stringari-03}) as well as for
classical nonlinear optics and plasmas~\cite{Berge-98}. For the linear Schr\"odinger
Equation with a random potential in one dimension all the states
are localized~\cite{Kramer-MacKinnon-93,
*Sheng-06}, as a manifestation of
Anderson localization~\cite{Anderson-58}. It is natural to ask what
is the asymptotic behavior for the NLSE with a random potential,
namely what is the outcome of the competition between Anderson
localization and nonlinearity. This question is directly
related to several experimental situations. For Bose-Einstein
Condensates  the nonlinear term in the NLSE models the
interaction between the atoms while the random potential may be
generated by lasers as in recent experiments~\cite{Schulte-05,*Fort-05,*Billy_et_al-08}.
In nonlinear optics the NLSE adequately describes
light propagation in a nonlinear medium where the randomness can be written with light that passes trough
a diffuser~\cite{Schwartz-07} or by manufacturing~\cite{Lahini-08}.

 Here the exploration will be performed in the framework
of a discrete one dimensional model (defined by (\ref{eq1})). The
specific elementary question in this field is: will an initially
localized wave packet spread to infinity in the long time limit?
Extensive numerical simulations~\cite{Pikovsky-Shepelyansky-08,*Flach-Krimer-Skokos-09,*Laptyeva-etal-10,*Mulansky-Pikovsky-10} (where although the full control of errors is impossible, one argues that the average results are
statistically meaningful)
exhibit sub-diffusion, with the width of the wave packet
growing in time as $t^{\sigma}$ with $\sigma\approx 1/6$.
On the other hand, recently
 it was argued that eventually the spreading should stop and the dynamics is eventually almost-periodic on a kind of KAM torus~\cite{Johansson-Kopidakis-Aubry-10}.
Rigorous studies~\cite{Wang-Zhang-09} lead to the conjecture that in the strong
disorder limit the spreading is at most logarithmic in time,
excluding sub-diffusion as the asymptotic behavior. Non-rigorous
results based on perturbation theory extend this conjecture beyond
the regime of strong disorder with the help of a bound on the
remainder term of the perturbation series~\cite{Krivolapov-Fishman-Soffer-10}, but  the  times available here at orders calculated so far
 turned out to be short compared with
numerical calculations where sub-diffusion was found. A major
difficulty in the exploration of this problem is that we do not
know how far in space and time one should go so that the result
can be considered {\em asymptotic}. One reason this problem is
complicated is the fact that during the spreading the effective number of
degrees of freedom increases enhancing chaos, but their amplitude
decreases suppressing chaos. This motivated the present work which
is designed to decide which of these competing effects wins. To
address this issue we develop here a scaling theory of weak chaos
in disordered nonlinear lattices, expecting it will be useful for
extending results beyond our computational ability. Scaling approaches proved to be extremely powerful in equilibrium and nonequilibrium statistical physics~\cite{Cardy-96,*Barabasi-Stanley-95}, and have been also very successful
in understanding Anderson localization~\cite{
Abrahams-etal-79,
*Deych-Lisyansky-Altshuler-00,
*Evers-Mirlin-08}. In this letter a scaling theory for the probability
distribution to observe chaos or regularity, based on the
computation of the largest Lyapunov exponent is developed and
tested for relatively small systems which are within our numerical
power. Such a  theory is expected to have
 predictive power when
extended to infinite size.

We study a nonlinear disordered medium described by the Discrete
Anderson Nonlinear Schr\"odinger Equation (DANSE) model for a complex field $\psi_n(t)$:
\begin{equation}
i \frac{d {\psi}_{n}}{d {t}} =\epsilon_{n}{\psi}_{n}
+J(\psi_{n+1}+ \psi_{n-1}) +{\mid{\psi_{n}}\mid}^2 \psi_{n} \;.
\label{eq1}
\end{equation}
The hopping is $J=(1+W)^{-1}$ while
$\epsilon_n$ are independent identically uniformly distributed in
$(-J W,J W)$.  With this rescaling the
eigenvalues $E$ of the linear part satisfy $|E|\leq 1+(1+W)^{-1}$.
Hence for strong disorder (large $W$, that is the focus of our
present work), the energies of the corresponding linear equation
are practically in the interval $(-1,1)$. Measuring the length
scale of the eigenstates of the linear problem $\mu$
by the inverse participation number $\mu^{-1}=\sum_k |\Psi_k|^4$,
we find that $\mu\approx1+W^{-1}$ for large $W$. By scaling the
amplitude of the field we set the coefficient of the nonlinear
term in (\ref{eq1}) to one. We consider DANSE model on a lattice
(ring) of length $L$, with periodic boundary conditions. While
DANSE model has two integrals of motion, the norm of the field
$N=\sum_{n} |\psi_n|^2$ and the total energy, only the norm is
important here; and in our treatment we do not control the energy which
is always chosen to be close to zero (center of the band of the
corresponding linear system).

Our goal in this letter is to study properties  of the dynamics as
a function of all relevant parameters: disorder $W$, norm $N$, and
lattice length $L$. Equivalently, we introduce the density
$\rho=N/L$ and consider the dependence on two intensive parameters
$W$ and $\rho$ and on the extensive parameter $L$.

\begin{figure}[tbh]
   \centering
\includegraphics[width=0.42\textwidth,angle=0]{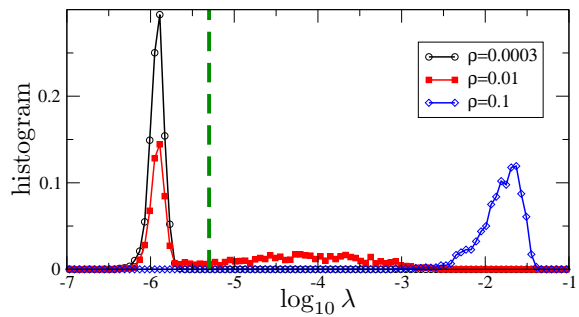}
\caption{Distributions of the largest Lyapunov exponent for $L=16$ and $W=10$, and three values of the field density. For $\rho=0.0003$ all the realizations are regular, for $\rho=0.1$ all are chaotic, but for $\rho=0.01$ a part of realizations are chaotic. Vertical dashed line indicates the border $\lambda=5\cdot 10^{-6}$.
\label{fig_le}
}
\end{figure}

We characterize the dynamics by means of the largest Lyapunov
exponent.  For a particular realization of disorder
$\{\epsilon_n\}$, we followed a dynamical solution of
Eq.~(\ref{eq1}) starting with a {\em uniform initial field}, and
for this trajectory calculated the largest Lyapunov exponent
$\lambda$ by a standard method. This calculation was repeated for
a large number of realizations of disorder, and the parameters
above. As a result, we can construct a distribution of largest
Lyapunov exponents over the realizations of disorder, for given
macroscopic parameters $W,\rho,L$. Several examples of these
distributions are presented in Fig.~\ref{fig_le}. We can see here
that for very small densities $\rho$, where the nonlinearity in
DANSE is very small, all the Lyapunov exponents are small and
close to $\approx 10^{-6}$. For a regular dynamics the largest
Lyapunov exponent should be exactly zero, but numerically, with a
finite integrating time ($T_{max}=10^6$ in our numerical
simulation; control runs with $T_{max}=10^7$ showed no significant difference) such a small value essentially indicates regular
(quasiperiodic) dynamics of the field. Contrary to this, for
large densities we observe Lyapunov exponents in the range $0.01-0.1$, what indicates for chaotic dynamics. For intermediate
densities we see that for some realizations of disorder the
dynamics is regular ($\lambda\approx 10^{-6}$), while for
other realizations much larger exponents are observed indicating
weak chaos. As we want to perform a statistical analysis rather
than going into details of particular dynamics for particular
realizations of disorder, we adopt the following operational
definition to distinguish between regularity and chaos: we
attribute all runs with $ \lambda<5\cdot 10^{-6}$ as regular ones,
and all runs with $ \lambda>5\cdot 10^{-6}$ as chaotic ones. In
this way we define the quantity of our main interest in this
letter, a probability to observe regular dynamics $P$.

\begin{figure}[tbh]
   \centering
\includegraphics[width=0.48\textwidth]{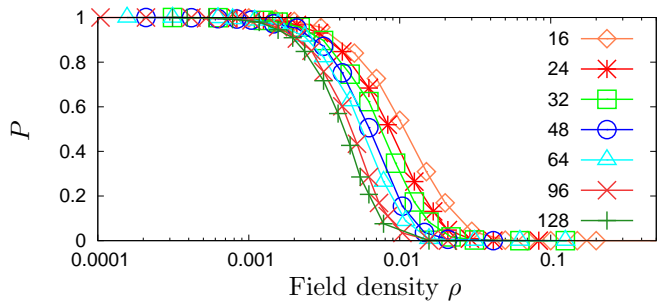}
\caption{The values of $P(\rho,W,L)$ for $W=10$ and different $L$, as functions of $\rho$. For the same density $\rho$ the probability to observe regularity decreases with $L$.
\label{fig_rprof}
}
\end{figure}

As  explained above, we have determined the probability of regular
dynamics $P(\rho,W,L)$ as a function of parameters of the model.
Typical profiles of $P$ for fixed disorder $W$ and different
values of $\rho$ and $L$ are depicted in Fig.~\ref{fig_rprof}. One
can see a typical sigmoidal function with limits $P\to 1$ for
$\rho\to 0$ and $P\to 0$ for $\rho\to\infty$.

\begin{figure}[tbh]
   \centering
\includegraphics[height=4.94cm]{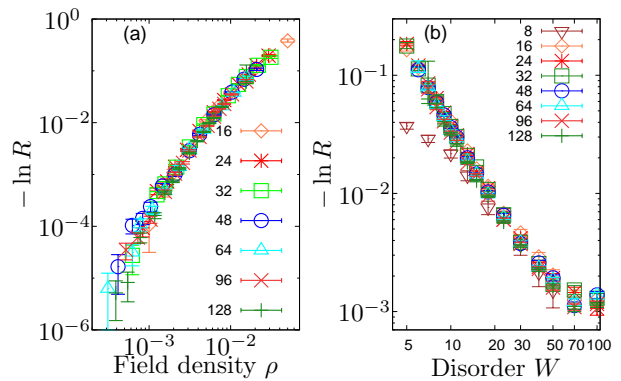}
\caption{Rescaled according to (\ref{eq_scl2}) probabilities to observe regular behavior, for different lattice sizes. (a): fixed disorder $W=10$, dependence on density (the same data as in Fig.~\ref{fig_rprof}); (b): fixed density $\rho=0.01$, dependence on the disorder.
\label{fig_rscall}
}
\end{figure}

We first concentrate on the dependence on the extensive parameter
$L$. As one can see from Fig.~\ref{fig_rprof}, for the fixed $W$
and $\rho$, the probability to observe regularity decreases with
the length $L$. This behavior can be understood as follows.
Suppose we fix $W,\rho$ and consider a lattice of large length
$L$. Let us divide this lattice into (still large) subsystems of
lengths $L_0$. How the probability to observe regularity on the large
lattice $P(\rho,W,L)$ is related to the corresponding
probabilities for smaller lattices $P(\rho,W,L_0)$? It is
reasonable to assume that to observe regularity in the whole
lattice we need to have all the subsystems regular, because any
one chaotic subsystem will destroy regularity. This immediately
leads us to the relation
\begin{equation}
P(\rho,W,L)=[P(\rho,W,L_0)]^{L/L_0}\;. \label{eq_scl}
\end{equation}
Equation (\ref{eq_scl})  implicitly assumes that chaos appears not
due to an interaction between the subsystems, but in each
subsystem (of length $L_0$) separately. This appears reasonable if
the interaction between the subsystems is small, i.e. if their
lengths are large compared to the length scale associated with
localization in the linear problem: $L_0\gg\mu$. On these scales
the various subsystems are statistically independent. This is the
content of (\ref{eq_scl}). It motivates the definition of the
$L$-independent quantity:
\begin{equation}
R(\rho,W)=[P(\rho,W,L)]^{1/L}\;. \label{eq_scl2}
\end{equation}
We check the scaling relation (\ref{eq_scl},\ref{eq_scl2}) in
Figs.~\ref{fig_rscall} and see that the data for lattices of sizes
$16<L<128$ collapse, so that  $R$ is independent of $L$.
Remarkably, a short lattice with $L=8$ obeys the scaling for large
disorders but deviates significantly for small disorders
$W\lesssim 10$; this corresponds to the expected validity
condition that $L$ should be larger than $\mu$ (the spatial size of
eigenfunctions).

\begin{figure}[tbh]
   \centering
\includegraphics[width=0.42\textwidth]{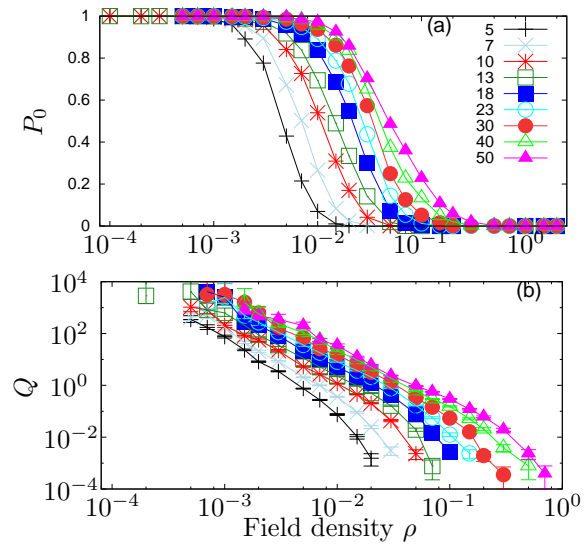}
\caption{Profiles of $P_0(\rho,W)$ vs $\rho$ for different values of $W$ (a) and the same data in terms of $Q(\rho,W)$ (b).
\label{fig_r161}
}
\end{figure}

The scaling relation (\ref{eq_scl}) describes dependence on the extensive parameter $L$ (and will allow us to extrapolate results to long lattices beyond our
numerical resources),
 so we can concentrate on considering dependencies on intensive parameters $W$ and $\rho$.
 Therefore below we fix $L=L_0=16$ and study the scaling properties of $P_0(\rho,W)=P(\rho,W,L_0)$.
 As this quantity is roughly a sigmoidal function of $\rho$ (see Fig.~\ref{fig_r161}a), it is
 convenient to perform a transformation to a new quantity $Q(\rho,W)$ as
$Q=\frac{P_0}{1-P_0}$.
%=\exp[2\text{arctanh}(2P_0-1)]
In this representation
\begin{equation}
P_0=\frac{Q(\rho,W)}{1+ Q(\rho,W)}=\frac{1}{1+Q^{-1}(\rho,W)}
\label{eq:r}
\end{equation}
so that the asymptotic behaviors $P_0\to 1$ as $Q\to\infty$ and
$P_0\to 0$ as $Q\to 0$ can be easily visualized as in
Fig.~\ref{fig_r161}b.

\begin{figure}[tbh]
   \centering
\includegraphics[width=0.46\textwidth]{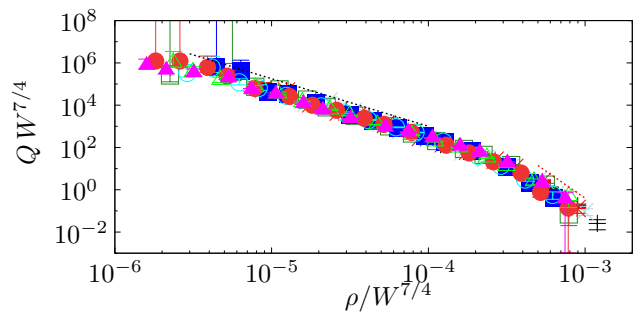}
\caption{(Color online) The same data as in Fig.~\ref{fig_r161} but in scaled coordinates. The black dashed and red dotted lines, showing asymptotics for small and large arguments of $q$, have slopes $\zeta=9/4$ and $\eta=5.2$.
\label{fig_r162}
}
\end{figure}

The next crucial observation is that the function $Q(\rho,W)$ is not an arbitrary function of $\rho$
and  $W$, but it can be written in a scaling form
\begin{equation}
Q=\frac{1}{W^\alpha}q\left(\frac{\rho}{W^{\beta}}\right)
\label{eq:sc1}
\end{equation}
where $q(x)$ is as usual a singular function at its limits
$q(x)\sim c_1 x^{-\zeta}$ for small $x$, while $q(x)\sim
c_2x^{-\eta}$ for large $x$. These are found from the straight
lines in Fig.~\ref{fig_r162}. The data of
Fig.~\ref{fig_r161}  collapses to one curve
as shown in Fig.~\ref{fig_r162}. This is the numerical
justification for (\ref{eq:sc1}). It also provides the values of
the exponents $\alpha=\beta=1.75$, $\zeta\approx 9/4=2.25$,
$\eta\approx 5.2$, $c_1\approx 2.5\cdot 10^{-7}$ and $c_2\approx1.8\cdot 10^{-18}$.

%allows us to collapse the data of Fig.~\ref{fig_r161} on a single
%curve, as demonstrated in Fig.~\ref{fig_r162}.

%\textbf{
The existence of the scaling function enables us to analyze the
behavior at various limits. Most interesting is the limit of small densities
where $Q$ is large (see Fig.~\ref{fig_r162}) and probability of regular behavior is close to one; correspondingly, the probability $P_{ch}$ to observe chaos is small. Using the general relation following from (\ref{eq_scl},\ref{eq:r},\ref{eq:sc1})
%}
\[
P(\rho,W,L)=\left[1+W^\alpha q^{-1}\left(\frac{\rho}{W^{\alpha}}\right)\right]^{-\frac{L}{L_0}}
\]
%\textbf{
we obtain
%}
\[
P_{ch}\approx -\log P= L L_0^{-1} W^\alpha q^{-1}\left(\frac{\rho}{W^{\alpha}}\right)
\]
%\textbf{
Using the asymptotics $q(x)\sim c_1 x^{-\zeta}$ we get\footnote{Related estimates have been discussed by D. Basko, arXiv:1005.5033v1 [cond-mat.dis-nn] (2010).}
%}
\begin{equation}
P_{ch}\approx LL_0^{-1} W^{\alpha(1-\zeta)} \rho^\zeta c_1^{-1}
\label{eq_ras}
\end{equation}

%\begin{equation}
%\label{PCH}
%Q\approx c_1\rho^{-\zeta}W^{-(\alpha-\beta\zeta)},
%\end{equation}
%therefore for large $Q$ using (\ref{eq_scl}) and (\ref{eq:r}) one finds
%\begin{equation} P(\rho,W,L)=P(\rho,W,L_0)^{\frac{L}{L_0}}\sim
%(1-c_1^{-1} W^{(\alpha-\beta\zeta)}  \rho^{\zeta})^{\frac{L}{L_0}}
%\label{eq_ras}
%\end{equation}
%\textbf{
Now let us assume that we consider the states with the same fixed
norm $N$ on lattices of different length $L$. Then $\rho=N/L$ and
from Eq.~(\ref{eq_ras}) it follows
%}
\begin{equation}
P_{ch}\approx \frac{L^{1-\zeta}N^\zeta W^{\alpha(1-\zeta)} }{c_1 L_0}=\frac{L^{-5/4}N^{9/4}}{c_1 L_0 W^{35/16}} \label{eq_ras1}
\end{equation}
%\textbf{
This quantity, as expected, grows  with the norm $N$ and
decreases with the disorder $W$. We see that because $\zeta>1$,
probability to observe chaos in large lattices at fixed norm tends
to zero. This result may have implications for the problem of
spreading of an initially local wave packet in large lattices. In
this setup the norm of the field is conserved, and the effective
density decreases in course of the spreading. If one assumes that
the dynamics follows the scaling above (although we established it
for a special setup of lattices of finite lengths with periodic
boundary conditions), and if one assumes that chaos is
essential for spreading, then one concludes that the spreading
should eventually stop as the probability to observe chaos 
%(a
%necessary ingredient for spreading) 
eventually vanishes. To
estimate, according to arguments above, at which length we can
expect  the spreading to stop, we have to start with large
densities (where the probability to observe regular dynamics is
negligible) and to estimate, at what lattice size $L_{max}$ chaos
extincts. Assume that this happens when the probability to observe
chaos reaches some small level $D$. Substituting $P_{ch}\approx D$
in Eq.~(\ref{eq_ras1}), we obtain for the following estimate for
$L_{max}$:
%}
\[
L_{max}=N^{\frac{\zeta}{\zeta-1}}W^{-\alpha} (D L_0c_1)^{\frac{1}{1-\zeta}}
\]
%\textbf{
Assuming $DL_0\approx 1$ and substituting the constants found, we obtain
%}
\begin{equation}
L_{max}(N,W)\approx 2\cdot 10^5\cdot N^{9/5}W^{-7/4}
\label{eq_estl}
\end{equation}
%\textbf{
This is our estimation for the maximal spreading of a wave packet of norm $N$ in a lattice with disorder $W$. Taking values typical for the
numerical experiments~\cite{Pikovsky-Shepelyansky-08,*Flach-Krimer-Skokos-09,*Laptyeva-etal-10,*Mulansky-Pikovsky-10}, namely $N=1,\; 1\leq W\leq 10$, we obtain from (\ref{eq_estl}) $L_{max}$ in the range from $3\cdot 10^4$ to $2\cdot 10^5$.
%}  
This explains, why in the numerical experiments, where typically values $L\approx 100$ are achieved, no saturation of the power law spreading is observed.

In conclusion, we established a full scaling theory for the
probability to observe chaos in disordered nonlinear Schr\"odinger
lattice: the scaling with the extensive parameter -- lattice
length -- is given by (\ref{eq_scl2}), while the scaling
dependence on intensive parameters (disorder and density) is given
by Eq.~(\ref{eq:sc1}). The found scaling indices indicate, that
for long lattices with the same norm chaos extincts and regularity
prevails. Furthermore, we use 
the system presented here as a model
for the chaotic region of high density typically observed at the initial stage  of evolution in all numerical experiments~\cite{Pikovsky-Shepelyansky-08,*Flach-Krimer-Skokos-09,*Laptyeva-etal-10,*Mulansky-Pikovsky-10}. In this context the
scaling with the length allows us to estimate the maximal length
that could be reached by spreading from initially local chaotic
wave packet. Remarkably, the scaling relations established include
large/small constants, what explains previous observations of
energy spreading over extremely large time scales. Nevertheless,
by applying the scaling we were able to estimate the final stages
of evolution from the studies of relatively small lattices.

\acknowledgments
AP thanks Lewiner Institute for Theoretical Physics (Technion) for hospitality. This work was partly supported by the Israel Science Foundation (ISF), by the US-Israel Binational Science Foundation (BSF), by the Minerva Center of Nonlinear Physics of Complex Systems, by the Shlomo Kaplansky academic chair.
We thank S. Aubry, D. Shepelyansky, B. Shapiro, A. Iomin, M. Mulansky, and Y. Krivolapov for useful discussions.

%Merlin.mbs v4.21 2009-07-09.
%

% \bibliographystyle{apsrev}
%\bibliographystyle{prsty}
% \bibliography{nld-old,nld-current,%
% pap-ab,pap-ce,pap-fg,pap-hj,pap-kl,%
% pap-mn,pap-oq,pap-rs,pap-tz,%
% pik,books,n-stand}

\end{document}